\documentclass[conference]{IEEEtran}

\usepackage{mathtools}
\usepackage{multirow}
\usepackage[font=footnotesize]{subcaption}
\usepackage[font=footnotesize]{caption}
\usepackage{amsthm}
\usepackage{graphicx}
\usepackage{color}
\usepackage{epstopdf}
\usepackage{amsfonts}
\usepackage{cite}
\usepackage{soul}
\PassOptionsToPackage{breaklinks}{hyperref}
\usepackage{xurl}
\usepackage{hyperref}

\usepackage{amssymb}
\usepackage{amsmath}
\usepackage{algpseudocode,algorithm}
\usepackage{setspace}
\graphicspath{{Figures/}}

\DeclareMathSizes{10}{9}{7}{6}
\usepackage{lipsum}
\usepackage{comment}
\usepackage{titlesec}

\begin{document}

%
 \addtolength\abovedisplayskip{-0.35\baselineskip}%
 \addtolength\belowdisplayskip{-0.35\baselineskip}%
\titlespacing{\section}{1pt}{\parskip}{-\parskip}
\titlespacing{\subsection}{1pt}{\parskip}{-\parskip}
\titlespacing{\subsubsection}{1pt}{\parskip}{-\parskip}

\title{Rate-Convergence Tradeoff of Federated Learning over Wireless Channel \vspace{-0.6em}}

\author{\IEEEauthorblockN{Ayoob Salari\IEEEauthorrefmark{1},
		Mahyar Shirvanimoghaddam\IEEEauthorrefmark{1}\IEEEauthorrefmark{2}, Branka Vucetic\IEEEauthorrefmark{1},  Sarah Johnson\IEEEauthorrefmark{2}}
\IEEEauthorblockA{\IEEEauthorrefmark{1}School of Electrical and Information Engineering, The University of Sydney, NSW 2006, Australia \\
\IEEEauthorrefmark{2}School of Electrical Engineering and Computing, The University of Newcastle, NSW 2308, Australia \\
Emails: \{ayoob.salari, mahyar.shm, branka.vucetic\}@sydney.edu.au, sarah.johnson@newcastle.edu.au
}\vspace{-1.8em}}

\maketitle

\begin{abstract}
In this paper, we consider a federated learning problem over wireless channel that takes into account the coding rate and packet transmission errors. Communication channels are modelled as packet erasure channels (PEC), where the erasure probability is determined by the block length, code rate, and signal-to-noise ratio (SNR). To lessen the effect of packet erasure on the FL performance, we propose two schemes in which the central node (CN) reuses either the past local updates or the previous global parameters in case of packet erasure. We investigate the impact of coding rate on the convergence of federated learning (FL) for both short packet and long packet communications considering erroneous transmissions. Our simulation results shows that even one unit of memory has considerable impact on the performance of FL in erroneous communication.
\end{abstract}


\IEEEpeerreviewmaketitle

\section{Introduction}\label{introduction}
\IEEEPARstart{I}{nternet} of things (IoT)-enabled applications and services have grown in popularity due to their potentials to improve human lives \cite{guo2021enabling}. By the year 2023, there will be 14.7 billion connected devices, with IoT devices making up half of that total \cite{Cisco}. It is envisioned that the number of concurrent connections increase from one million per square kilometre in 5G to ten million per square kilometre in 6G \cite{mostafa2021aggregate,yuan2021noma}. Together with artificial intelligence (AI), the next generation of wireless technologies intends to enable the communication infrastructure required for the massive IoT (mIoT) use cases \cite{lin2021ai,prasad2021advance}. In mIoT,  a large number of low-cost low-power endpoints communicate with the central node (CN) \cite{vaezi2022cellular}. Conventionally, users send their local data to a computationally capable central controller for the purpose of training a deep network model. However, such an approach is impractical in a mIoT scenarios owing to concerns about privacy, limited power, and often inadequate communication bandwidth, all of which impose a significant strain on the communication links \cite{imteaj2021survey}. Federated learning (FL) is an alternative approach, in which each user trains a local model based on its own data and the global parameter and transmits the updated model parameter to the CN. The CN adjusts the global model parameter with a weighted average of users' updates and broadcasts the new global parameters to the network. This procedure is continued until convergence occurs \cite{chen2022federated}.

%
It is critical to comprehend the difficulties associated with FL. Since each device's dataset is acquired by the user itself, it depends on the client's local environment. As a result, not only the users' datasets are non-$\mathrm{i.i.d}$ throughout the network, but also the size of users' datasets may vary significantly \cite{sattler2019robust}. This statistical heterogeneity has an impact on the convergence mechanism of FL and lowers model accuracy. One can use an adaptive averaging strategy or apply data sharing mechanism to reduce the impact of non-$\mathrm{i.i.d}$ datasets \cite{zhao2021federated}. 
To cope with the heterogeneity of systems, FL must handle a variety of devices that have differing amounts of memory and processing power as well as differing battery sizes and storage capacities \cite{li2020federated}. Weight-based federated averaging is considered as a solution to this problem \cite{nguyen2020efficient}.
Another challenge of FL is the tradeoff between devices processing power and the communication overhead. Local processing power is substantially faster than communication in the network. On the one hand, increasing the number of local computing iterations leads in a decrease in the number of network communications. On the other hand, since users are constrained in terms of power, increasing the number of local iterations depletes the battery and the device's ability to communicate with the CN is severely limited \cite{wang2019adaptive}.

While substantial effort has been made to tackle the aforementioned FL challenges, it was presumed that one could simply use FL in wireless networks and devices could communicate with CN without error \cite{sattler2019robust,zhao2021federated, li2020federated,nguyen2020efficient,wang2019adaptive,amiri2021convergence}. However, in a mIoT configuration, where massive power and bandwidth limited devices with low computing capacity transmit short packet data to a CN, the wireless channels are unreliable and communication is often erroneous \cite{NOMA2021Yue}. Despite recent investigations of FL algorithms over wireless fading channels, they are constrained in a number of ways \cite{amiri2020federated,zhang2020federated, chen2020joint}. The study in \cite{zhang2020federated} only indicated a drop in model accuracy in the event of a communication failure. To alleviate the impact of fading channels, the authors in \cite{amiri2020federated} advocated one device transmission in each iteration; however, in a mIoT scenario where a large number of users desire to connect with CN, this strategy is unsuitable. The authors in \cite{chen2020joint} presented resource allocation for minimizing model loss in case of communication error. However, they neglected the network's statistical heterogeneity, which has a significant effect on the system performance. Furthermore, the accuracy of their suggested model is less than 90\%. In \cite{Shirvanimoghaddam2022Federated}, the authors investigated the FL algorithm over packet erasure channels and demonstrated that when CN depends solely on fresh local updates, both loss and accuracy of the model fluctuate. However, it is yet unknown how the physical layer parameters such as the code rate and blocklength affect the convergence of FL in erroneous communication.

In this paper, we provide a realistic implementation of FL in a mIoT scenario that takes both the coding rate and the packet transmission error into consideration. We model the communication channel as a packet erasure channel, in which the CN either successfully receives local updates from devices or, with a certain probability, the packet is erased. To mitigate the impact of erasure on FL convergence, we assume that the CN stores model parameters in memory. We investigate two different schemes; 1) CN has adequate memory to record the last local update of each user, and 2) CN has limited capacity and caches the previous $m$ global updates. We analyse two distinct communication scenarios, i.e., short packet and long packet communications. The erasure probability is determined by the blocklength, code rate, and SNR. We show that for a fixed transmission power, lowering the code rate reduces the probability of error while increasing the reliability. However, because of the low code rate, there will be a greater amount of communications between the CN and devices. Therefore, there is a trade-off between code rate, convergence time, and FL model accuracy.

The remainder of the paper is structured as follows. Section II describes the system model and the various FL techniques used in erroneous communications. Section III analyzes the performance of FL algorithm with erroneous communications. Section IV contains numerical findings. Finally, Section V concludes the paper.

%
%
%
\section{System Model}  \label{System Model}
We consider a federated learning system comprising of one central node (CN) and a set $\mathcal{U}$ of $U$ IoT devices with local datasets, $\mathcal{D}_1$, $\mathcal{D}_2$, $\dots$, $\mathcal{D}_U$. Dataset of each device $u$ is defined as $\mathcal{D}_u = \{(x_u^1, y_u^1), (x_u^2, y_u^2), \cdots, (x_u^{D_u}, y_u^{D_u}) \}$ where $D_u$ is the dataset size of user $u$. The total amount of training data stored by all users is stated as  $D=\sum_{u=1}^{U} D_u$. The loss function, which varies depending on the learning model, is used to evaluate the FL algorithm's performance.  For a linear regression learning model, the loss function can be written as  $f(\boldsymbol{\omega},\boldsymbol{x},y) = \frac{1}{2} \parallel y - \boldsymbol{\omega}^T \boldsymbol{x} \parallel^2$, while in the case of neural network it is $f(\boldsymbol{\omega},\boldsymbol{x},y) = \frac{1}{2} \parallel y - f_{nn}(\boldsymbol{x};\boldsymbol{\omega})  \parallel^2$, where $f_{nn}(\boldsymbol{x};\boldsymbol{\omega}) $ is the learning output of neural network. 

Suppose IoT devices wish to send their local updates in messages with a length of $k$ bits to the CN. An encoder is used to map these messages to codewords of length $n$ using a channel code of rate $R=k/n$. The received signal of user $u$ at the CN is 
\begin{equation} \label{eq:1}
    r_u = h_u z_u  + n,
\end{equation}
where $r_u$ is the received signal, $z_u$ is the transmitted signal of user $u$, and $n$ is zero-mean additive white Gaussian noise (AWGN) with variance $\sigma^2$ and $h_u$ is the channel gain that follows a zero-mean circularly symmetric Gaussian distribution, i.e., $h_u\sim\mathcal{CN}(0,1)$. The channel is block fading, which is constant over each packet duration and independently varies across packets. Assuming the user power as $P$, the received signal-to-noise (SNR) at CN is $\gamma_u = \gamma_0 ||h_u||^2$, where $\gamma_0 = P/\sigma^2$. 
%

Packet errors will arise as a consequence of block fading and channel noise. Here, we further simplify the communication channel as a packet erasure channel in which devices' packets are either erased with a probability of $\epsilon$ or successfully received at the CN with a probability of  $1-\epsilon$. We examine two distinct scenarios of short and long packet communication. In the case of long packet transmission, the packet erasure rate can be accurately estimated as follows:
\begin{align}
\epsilon = \text{Pr}(\gamma < \gamma_{th})
 \label{error long packet} 
\end{align}
where $\gamma_{th} = 2^R - 1$. 
For the short packet communication, authors of \cite{polyanskiy2010channel} showed that the packet error rate at the receiver can be written as 
\begin{align}
\epsilon \approx Q \left(\frac{n C(\gamma) - k + 0.5 \log_2(n)}{\sqrt{n V(\gamma)}}   \right) \label{error short packet} 
\end{align}
where  $C(\gamma) = \log_2(1+\gamma)$ is the channel capacity, $V(\gamma) = \log_2^2{(e)} \left( 1- {(1+\gamma)^{-2}}  \right)$ is the channel dispersion, and $Q(.)$ is the standard $Q$-function \cite{polyanskiy2010channel}. 

While most uplink transmissions are erroneous, we assume CN utilizes the whole spectrum and transmits with high power on the downlink, resulting in error-free broadcasts.

%
%
\subsection{Principles of FL in a error-free transmission}
In each iteration of FL, each device $u$ computes its local update $\boldsymbol{\omega}_u$ (usually using a few epochs of gradient descent), and transmit their updated parameter of the trained model to the CN. Next, the CN calculates the average weight, $\boldsymbol{\omega}$, by aggregating all the local updates. Then, CN broadcasts the updated global parameter to be used by devices for the next iteration of FL. 

Let us consider the loss function of device $u$, which calculates the model error on its data set $\mathcal{D}_u$ as 
\begin{equation}
   F_u(\boldsymbol{\omega}) = \frac{1}{D_u} \sum_{i=1}^{D_u}f(\boldsymbol{\omega},\boldsymbol{x}_u^i,y_u^i) 
\end{equation}
Employing the gradient descent (GD) approach, the local parameter of device $u$ at time $t$ can be computed as 
\begin{equation}
   \boldsymbol{\omega}^{(t)}_u = \boldsymbol{\omega}^{(t-1)} - \eta \nabla F_u(\boldsymbol{\omega}^{(t-1)})
\end{equation}
where $\eta$ is the learning rate. Once the IoT device has computed its own local parameter, it will transmit the updated parameter to the CN through an error-free channel, and the CN will aggregate all of the received local parameters to compute the global update using
\begin{equation}
   \boldsymbol{\omega}^{(t)} = \frac{1}{D} \sum_{u=1}^{U} D_u \boldsymbol{\omega}_u^{(t)}.
\end{equation}
After calculating the global parameter, CN will broadcast it throughout the network. One could combine the last two steps of FL and calculate the global update as
\begin{equation}
   \boldsymbol{\omega}^{(t)} = \boldsymbol{\omega}^{(t-1)} - \frac{\eta}{D} \sum_{u=1}^{U} D_u \nabla F_u(\boldsymbol{\omega}^{(t-1)})
\end{equation}
%
\subsection{FL in erroneous communication}
We will discuss erroneous channels between IoT devices and CN in this section. 
Given the SNR, $\gamma$, blocklength, $n$, and code rate, $R$, the probability that CN does not receive the local parameters can be calculated.
\subsubsection{FL in erroneous communication without CN memory}
The number of local parameters received by the CN during each communication round may vary depending on the channel quality. In general, the global parameter can be calculated using
\begin{align}
     \boldsymbol{\omega}^{(t+1)} = \frac{\sum_{u=1}^U{I_u D_u \boldsymbol{\omega}_u^{(t)}}}{\sum_{u=1}^U{I_u D_u}},
\end{align}
where $I_u \in \{0 , 1\}$ is the parameter that shows if CN receives the local update of user $u$ correctly, which follows Bernoulli distribution: 
  \[
    I_u = \left\{\begin{array}{ll}
        1; & \text{with probability } 1-\epsilon_u(\gamma,n,R), \\
        0; & \text{with probability } \epsilon_u(\gamma,n,R).
        \end{array}\right.
  \]
In this scenario, the number of local updates at CN may vary across iterations.
%
\subsubsection{FL in erroneous communication with CN memory} \label{CN with memory}
\begin{figure}[!t]
     \centering
     \begin{subfigure}[t]{0.69\columnwidth}
         \centering
         \includegraphics[width=0.95\columnwidth]{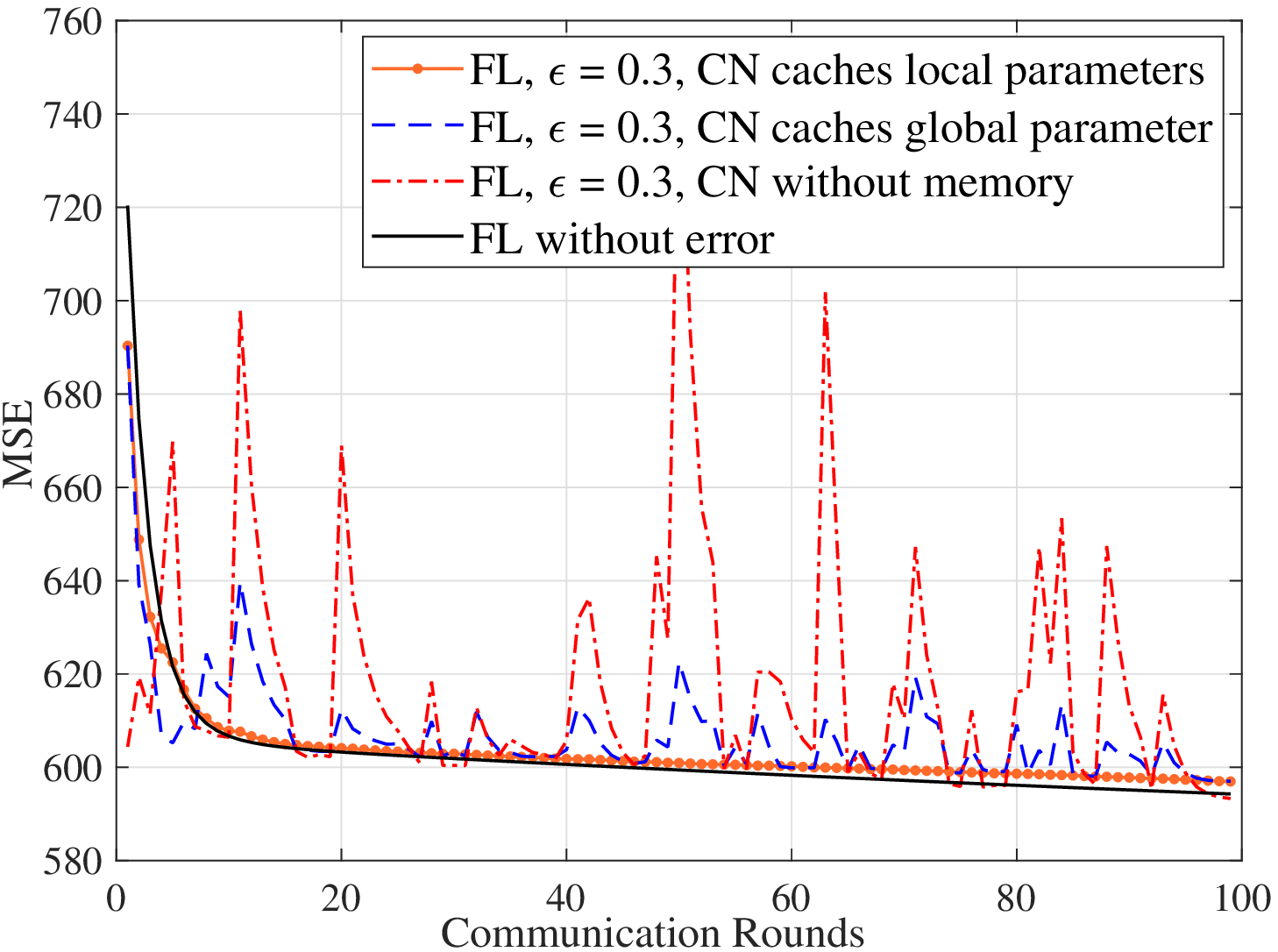}
         \caption{MSE vs. Communication round}
         \label{fig:F1a}
     \end{subfigure}
     \hfill
     \begin{subfigure}[t]{0.29\columnwidth}
         \centering
         \includegraphics[width=0.95\columnwidth]{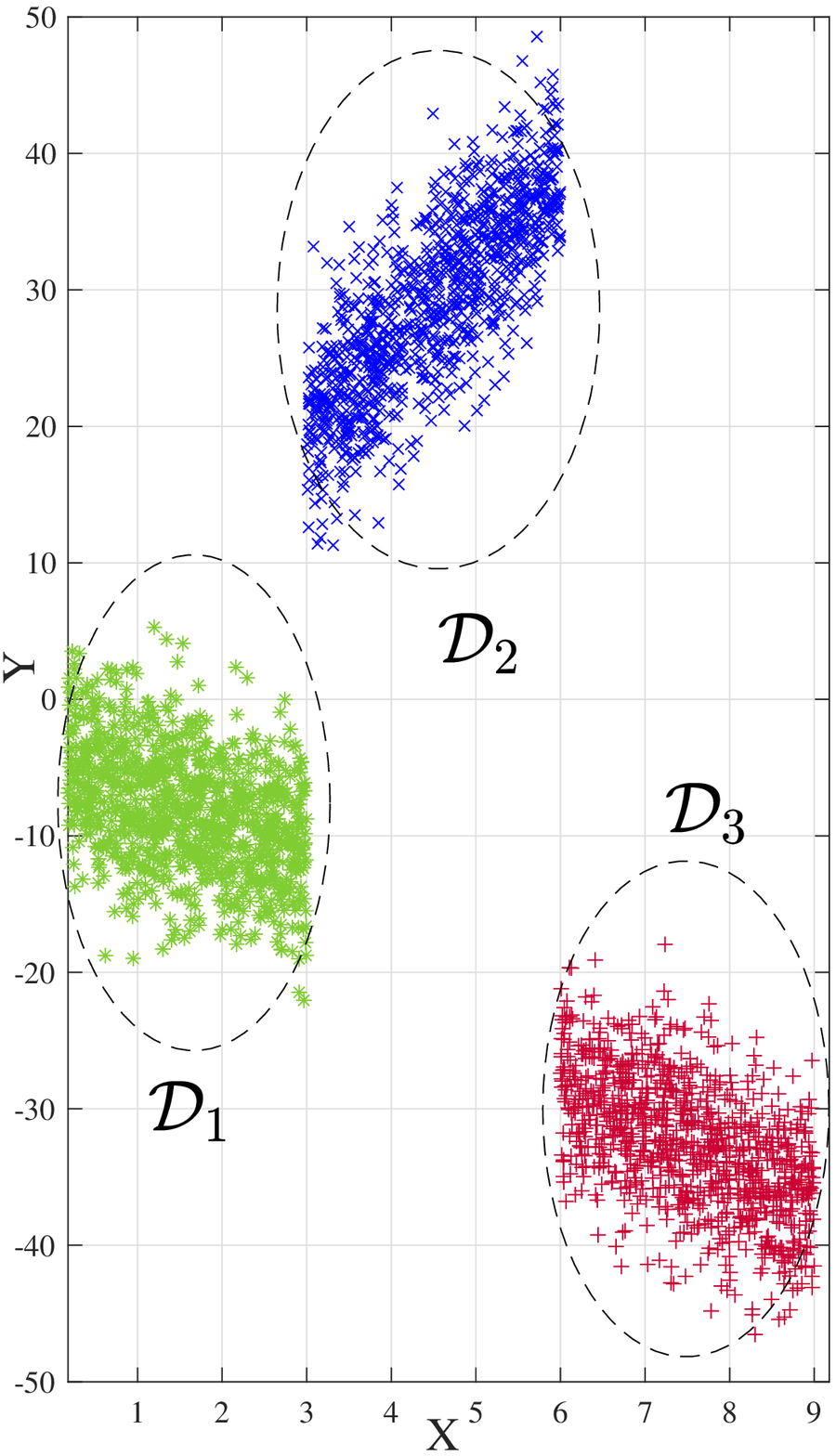}
         \caption{Non-$\mathrm{i.i.d.}$ Dataset}
         \label{fig:F1b}
     \end{subfigure}
        \caption{Impact of erasure channel on the convergence of FL, for $U=3$, $|D_u|=100$, $\eta = 0.05$ and 2 iterations at devices is used.}
        \label{fig:comparison}
    \vspace{-0.5em}
\end{figure}

We will assume in this part that CN has a memory to store model parameters.
\paragraph{CN caches user's local parameter} \label{CN caches users}
In this setup, we suppose the CN has a memory dedicated to storing each device's most recent local parameter. To compute the global parameter at each communication round, CN employs the fresh local parameter for users with successful transmissions and reuses the stored local parameter for users with an erroneous channel. Therefore, the number of users that participate in the update of global parameter would be fixed.
The global update can be computed as
\begin{align}
     \boldsymbol{\omega}^{(t+1)} = \frac{1}{D}\sum_{u=1}^U D_u \left(\boldsymbol{\omega}_u^{(t)} I_u +\boldsymbol{\omega}_u^{(t-1)} (1-I_u)\right).
\end{align}
%
\paragraph{CN caches global parameters} \label{CN caches global}
In an mIoT situation, where a large number of users transmit their local updates to the CN, it may be problematic for the CN to maintain a dedicated memory pool for storing all users' past local updates. In this case, we assume that CN has a finite amount of memory that it utilises to preserve the previous $m$ global updates. The global update can be calculated using:
\begin{equation}
     \small
     \boldsymbol{\omega}^{(t+1)} = \frac{1}{D}\sum_{u=1}^U D_u \left(\boldsymbol{\omega}_u^{(t)} I_u +(1-I_u) \sum_{i=0}^{m-1} \alpha_{t-i}  \boldsymbol{\omega}^{(t-i)}\right),
\end{equation}
where $\sum_{i=0}^{m-1} \alpha_{t-i} \boldsymbol{\omega}^{(t-i)}$ is the weighted average of last $m$ global updates, and $\alpha_{t-i}$ represents the weight of the global parameter at time instant $t-i$, where  $\sum_{i=0}^{m-1} \alpha_{t-i}  = 1 $. 
Fig. \ref{fig:comparison} compares erroneous and error-free communication for different FL schemes in short packet communication scenario. It can be noticed that the best performance is for the case that CN has memory to store the local parameters of IoT devices. When CN has a restricted memory space, instead of saving all local parameters of IoT devices, CN may save previous $m$ global updates (\ref{CN caches global}), however performance deteriorates in comparison to the scenario when CN memorises local values.
\begin{figure*}[t] 
\subfloat[$\gamma_0 = -3$ dB \label{fig:Local-different Rate-low SNR}]{%
\includegraphics[width=0.81\columnwidth]{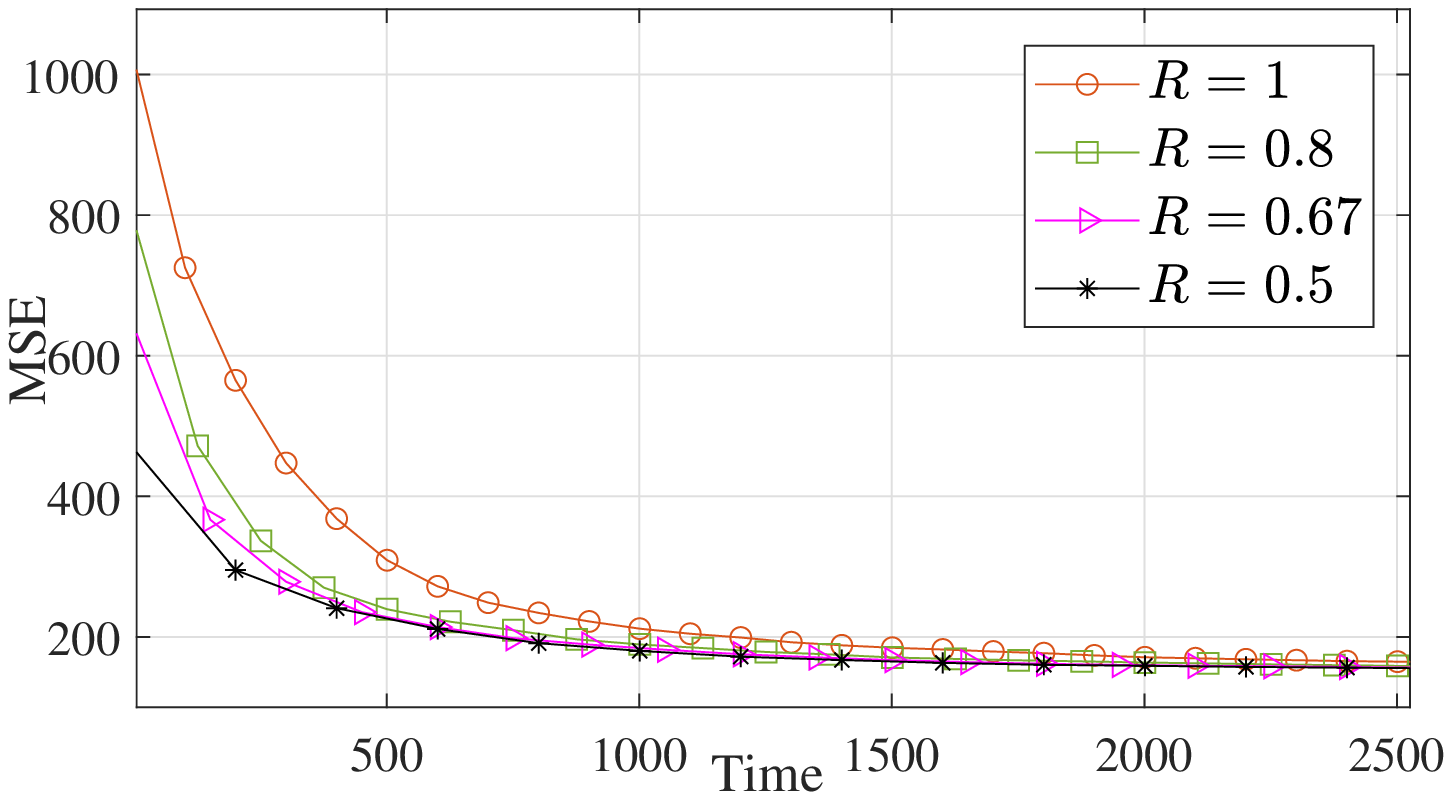}
}
\hfill
    \subfloat[$\gamma_0 =  3$ dB  \label{fig:Local-different Rate-high SNR}]{%
\includegraphics[width=0.81\columnwidth]{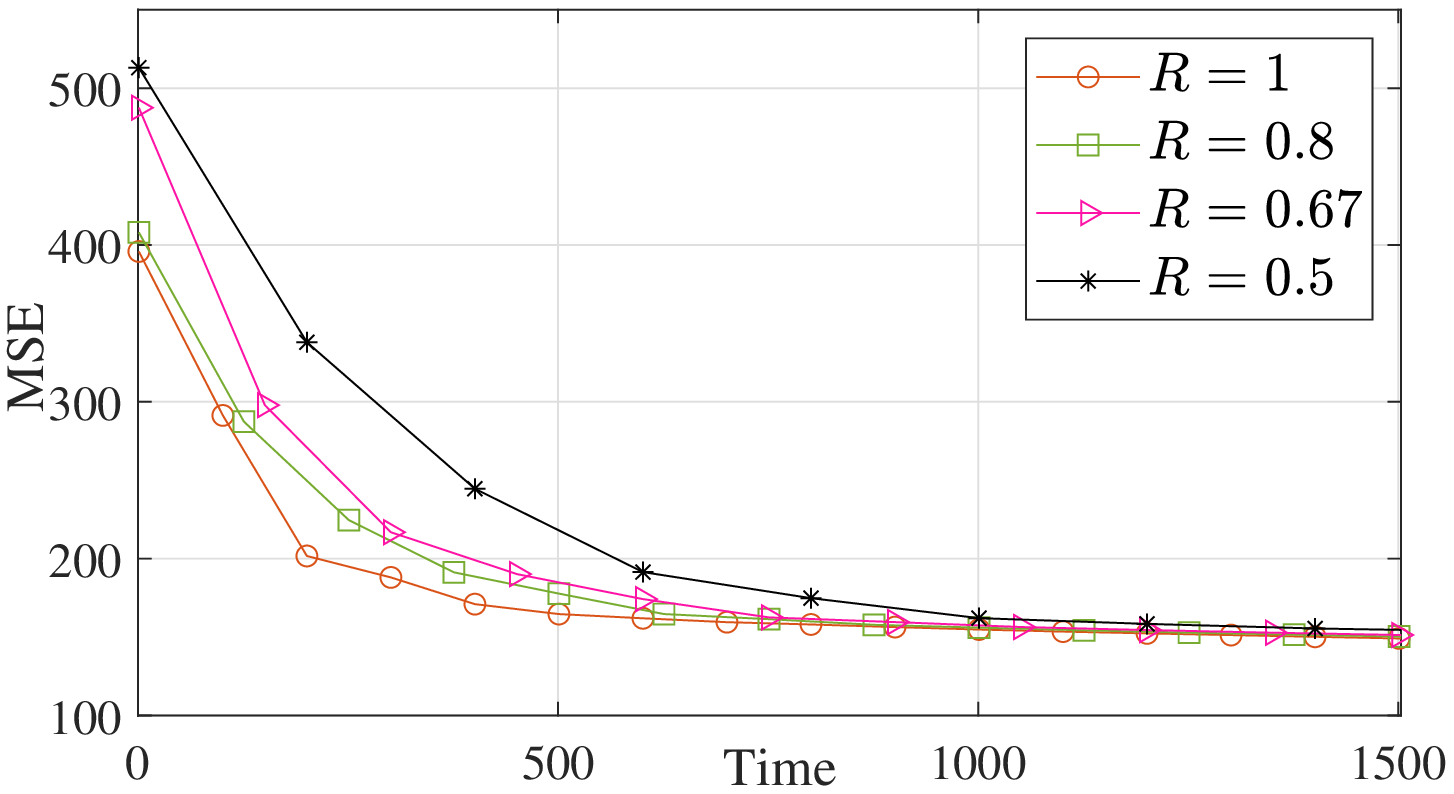}
}
\hfill
\subfloat[Non-$\mathrm{i.i.d.}$ Dataset \label{fig:Dataset-10 Users}]{%
\includegraphics[scale=0.5]{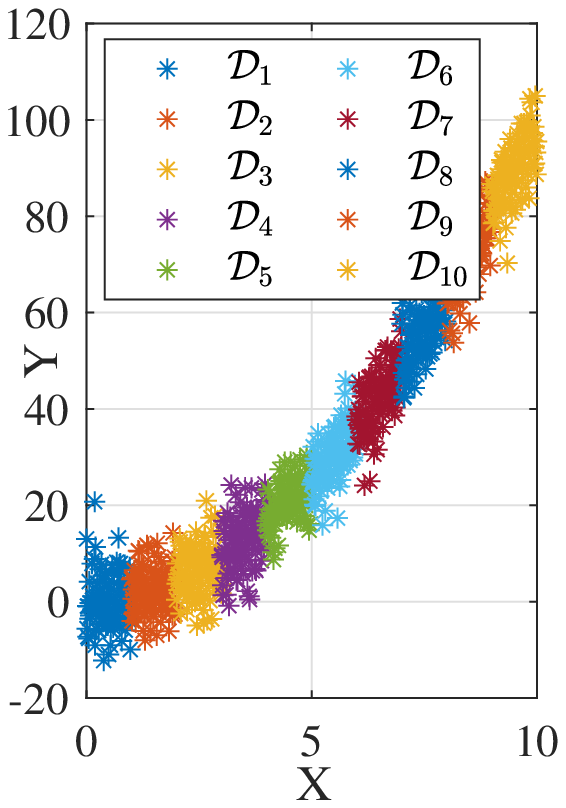}
}
\caption{\small{Impact of Rate and SNR on convergence of FL when CN stores local parameters of devices in erroneous communication, for $U=10$, $|D_u|=100$, $\eta = 0.05$ and 1 iteration of GD at device is applied.}}
\label{fig:Local - Rate vs Convergence}
\vspace{-1.5em}
\end{figure*}
%
\section{Performance Analysis}  
Without loss of generality, we assume that all devices have the same code rate and training dataset size, i.e., $R_u=R, D_u = D/U, \forall u \in \mathcal{U}$. Thus, all the devices have a similar number of packets to transmit. Subsequently, in an erroneous communication, where CN does not have memory, the global update can be written as 
\begin{align} \label{simplified No memory eq}
     \boldsymbol{\omega}^{(t+1)} = \frac{\sum_{u=1}^U{I_u \boldsymbol{\omega}_u^{(t)}}}{\sum_{u=1}^U{I_u }}.
\end{align}
Since $I_u$ follows a Bernoulli distribution, i.e., $I_u \sim \text{Ber}(1-\epsilon_u), \forall u \in \mathcal{U}$, and the channels are independent, the probability mass function $(\mathrm{pmf})$ of global parameter can be calculated as \cite{ayyala2020high,Shirvanimoghaddam2022Federated}
\begin{align} \label{pmf}
\mathrm{Pr}\left\{\boldsymbol{\omega}^{(t+1)} = \frac{\sum_{u=1}^U{I_u \boldsymbol{\omega}_u^{(t)}}}{\sum_{u=1}^U{I_u }}\right\} =\prod_{u=1}^U\epsilon_u^{1-I_u}(1-\epsilon_u)^{I_u}.
\end{align}
One can easily see that the random characteristic of erasure occurrences is critical to $(\ref{pmf})$, and results in fluctuation of the global parameter. Even though the instantaneous loss of the FL model in erroneous communication without memory fluctuates, the expected value of loss converges.

To further examine the distribution of global update in the case of erroneous communication without memory at CN, we analyse each component of $(\ref{simplified No memory eq})$. Although the denominator of $(\ref{simplified No memory eq})$ follows a Bionomial distribution, it is not straightforward to develop a general equation for the $\mathrm{pmf}$, unless the number of devices are very small, which is not the case for mIoT.

In \cite{le1960approximation}, Le Cam has proved that the sum of independent Bernoulli random variables, that are not necessarily identically distributed, where $\mathrm{Pr}(I_u = 1) = 1-\epsilon_u, \forall u \in \mathcal{U}$, has approximately a Poisson distribution with parameter $\lambda_m = (1-\epsilon_1) + (1-\epsilon_2) + \cdots + (1-\epsilon_m)$.
It has been shown that the sum of the absolute differences between the $\mathrm{pmf}$ of $Y=\sum_{u=1}^U{I_u }$ and the $\mathrm{pmf}$ of the Poisson distribution with parameter $\lambda_m$ is no more than twice the sum of the squares of the $(1-\epsilon_u), \forall u \in \mathcal{U}$, i.e.,
\begin{align} \label{Le Cam Eq}
\sum_{j=0}^\infty \left| \mathrm{Pr}(S_m=j) - \frac{\lambda_m^j e^{-\lambda_m}}{j!} \right| < 2\left( \sum_{u=1}^m  (1-\epsilon_u)^2 \right)
\end{align}
where $S_m = I_1 + \cdots + I_m$  follows a Poisson binomial distribution.
Larger $\lambda_m$, results in better approximation. This indicates that the more IoT devices who communicate with CN successfully, the more accurate the estimate for the sum of the Bernoulli random variables.  It's worth noting that the summation on the right hand side of equation $(\ref{Le Cam Eq})$ will not exceed  $9$ times the largest $(1-\epsilon_u), \forall u \in \mathcal{U}$ \cite{serfling1978some}.
While using Le Cam's equation, one can easily approximate the distribution of denominator of $(\ref{simplified No memory eq})$ at each communication round based on the number of IoT devices communicating with CN, the numerator has an entirely different tale. 
In \cite{daskalakis2015learning}, authors have studied the learnability of the weighted sum of independent Bernoulli random variables $Y = \sum_{u=1}^U{I_u \boldsymbol{\omega}_u}$. They developed an algorithm that given $U, \boldsymbol{\omega}_1, \cdots,\boldsymbol{\omega}_U $ and access to independent draws from $Y$, can find the required number of samples and running time needed to learn the weighted sum of independent Bernoulli random variables. However, the CN does not have the luxury of having access to this information. Therefore, it is not possible to learn the distribution of $(\ref{simplified No memory eq})$.

For the case of erroneous communication when CN has memory (\ref{CN with memory}), the global update parameter can be seen as the sum of two weighted sums of independent Bernoulli random variables. Let us consider $\mathcal{S}(t)$ as the set of devices that CN received their packet at time $t$ and $\mathcal{F}(t)$ as the set of devices that their packet have been erased, with size $|S(t)|$ and $|F(t)|$, respectively.
Hence, one can use the Hoeffding's inequality, which is an extension of Chernoff's bound for Bernoulli random variables, to establish an upper bound on the probability that global parameter deviates from its expected value by more than a certain amount. Using Chernoff-Hoeffding theorem, for any $\beta \in (0,1]$ we have
\begin{align} 
\text{Pr}\Big[ \boldsymbol{\omega}^{(t)} > (1+\beta) \frac{\kappa_1}{2} \Big] < \exp(-\beta^2 U \frac{\kappa_1}{6}) \label{Hoeffding Eq1} \\
\text{Pr}\Big[ \boldsymbol{\omega}^{(t)} < (1+\beta) \frac{\kappa_2}{2} \Big] < \exp(-\beta^2 U \frac{\kappa_1}{4}) \label{Hoeffding Eq2}
\end{align}
where $\kappa_1 = \text{min}\{ \frac{|S(t)|}{U} , \frac{|F(t)|}{U} \}$ and $\kappa_2 = \text{max}\{ \frac{|S(t)|}{U} , \frac{|F(t)|}{U} \}$ \cite{canonne2020survey}.
In \cite{Shirvanimoghaddam2022Federated}, it has been proved that for a convex and L-smooth loss function, the FL algorithm in erroneous communication, where CN caches local parameter of IoT devices (\ref{CN caches users}), converges to global optima.
In a mIoT configuration, it may be impractical to retain the prior local value of all IoT devices. Our simulation results shows that the FL algorithm in erroneous communication will converge provided just a few global parameters are stored (\ref{CN caches global}).
\begin{figure*}[!t] 
\subfloat[$\gamma_0 = -3$ dB \label{fig:Global-different Rate-low SNR}]{%
\includegraphics[width=1\columnwidth]{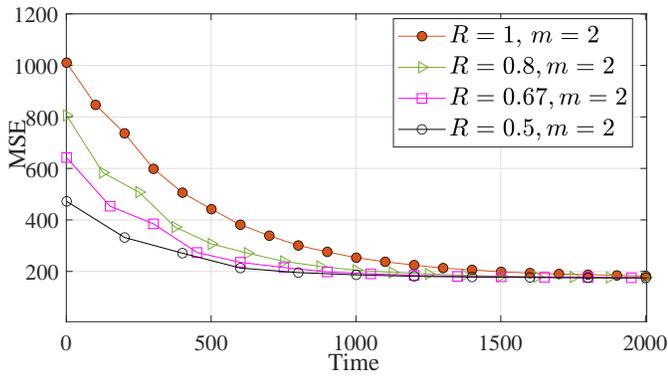}
}
\hfill
    \subfloat[$\gamma_0 =  3$ dB  \label{fig:Global-different Rate-high SNR}]{%
\includegraphics[width=1\columnwidth]{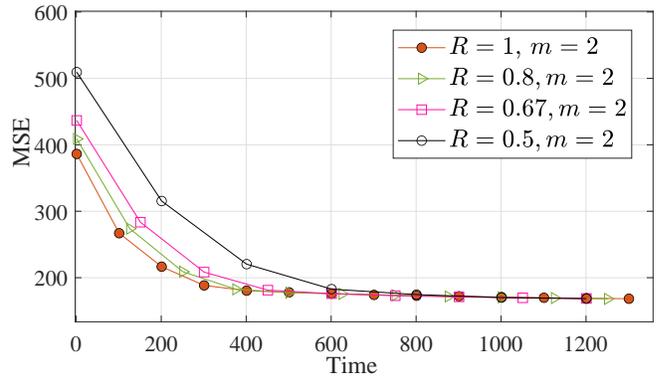}
}
\caption{\small{Impact of Rate and SNR on convergence of FL when CN store global parameters in erroneous communication, for $U=10$, $|D_u|=100$, $\eta = 0.05$, $m = 2$ and 1 iteration of GD at device is applied.}}
\label{fig:Global - Rate vs Convergence}
\vspace{-1 em}
\end{figure*}
\begin{figure*}[!t] 
\subfloat[CN stores global updates, $m = 2$ \label{fig:3D Global - short time}]{%
\includegraphics[width=1\columnwidth]{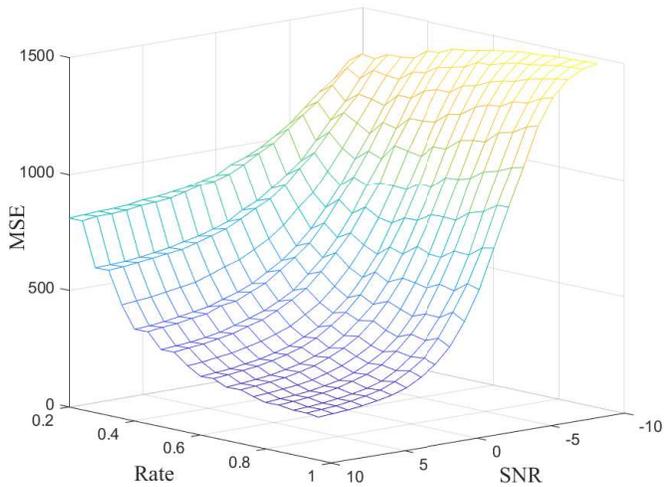}
}
\hfill
    \subfloat[CN stores local update of users  \label{fig:3D Local - short time}]{%
\includegraphics[width=1\columnwidth]{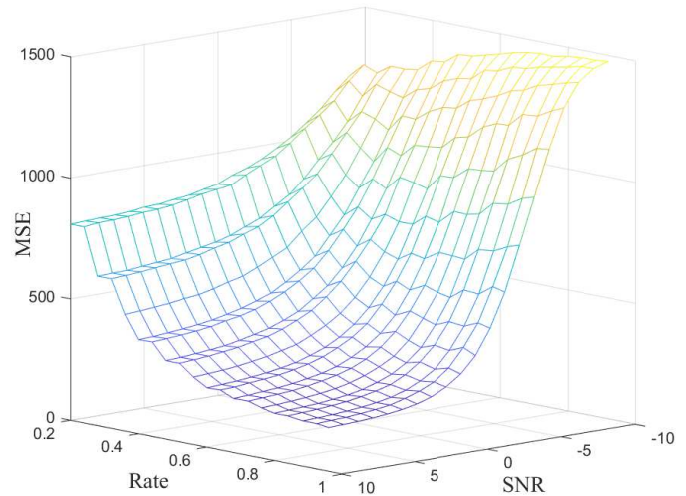}
}
\caption{\small{Rate vs. SNR vs. MSE of FL in erroneous communication, for $U=10$, $|D_u|=100$, $\eta = 0.005$, $\text{time} = 1500$ $T_s$ and 1 iteration of GD at device is applied.}}
\label{fig:3D - short time}
\vspace{-1.5 em}
\end{figure*}
%
%
%
\section{Numerical Results and Discussion}
We start by investigating the performance of FL in short packet communications, where we choose a message length of $k=100$ bits due to the fact that the normal approximation for AWGN channel is relatively good for packet lengths of $n \geq 100$ bits and $R \geq 0.5$ \cite{polyanskiy2010channel}. We consider there are $U=10$ users in the mIoT scenario. As shown in Fig. \ref{fig:Dataset-10 Users},  the non-$\mathrm{i.i.d.}$ datasets are created using non-linear model $y=x^2 + \lambda$, where $\lambda\sim\mathcal{N}(0,5)$. 

We consider the symbol duration as the time unit. We assume for all approaches the devices use the same modulation and that the symbol duration is the same. When the rate $R$ is reduced, the packet length increases and the time required to transmit the packet increases. The duration of each communication round is considered to be equivalent to the packet duration, i.e., $n$ symbols. Furthermore, Monte-Carlo simulations are used in short-packet communications, i.e. we ran 100 simulations and averaged the results.

Fig. \ref{fig:Local - Rate vs Convergence} and Fig.  \ref{fig:Global - Rate vs Convergence} show the global loss of FL based on the coding-rate $R$ in short packet erroneous communication, when CN stores local update of IoT devices and global updates, respectively. Considering fading channel, given $R$ and SNR, we use normal approximation bound (\ref{error short packet}) to calculate the probability of erroneous channel. Fig.\ref{fig:Local-different Rate-high SNR} and Fig. \ref{fig:Global-different Rate-high SNR} illustrate that in the high SNR regime, reducing the code rate results in lower convergence time. Because the users' power is sufficient to overcome the fading and noise, devices with higher rates will communicate more often than those with lower rates over a given learning period, leading the MSE to converge faster. However, in the low SNR regime, lower code rates are preferable for a shorter convergence time since we must compensate for the noisy channel by adding more parity bits to the message.

%
Fig. \ref{fig:3D - short time} shows the loss performance of erroneous short packet communication as a function of $R$ and SNR. We set the time to $1500$ $T_s$, where $T_s$ is the symbol duration. As can be seen, when the SNR is low, for a descent loss performance, we need lower $R$ to mitigate the impact of high noise level and fading. However, in high SNR scenarios, it is better to increase the code rate $R$. This is due to the fact that when $R$ is low, the packet length $n$ increases, and subsequently the number of communication rounds between devices and CN decreases and  leaving users with little time to achieve ideal performance.
Fig. \ref{fig:3D - short time} clearly demonstrates that when the $R$ grows, the number of communications for given times will increase and as a result, the MSE will decrease.

\begin{figure}[!t]
    \centering
    \includegraphics[width=1\columnwidth]{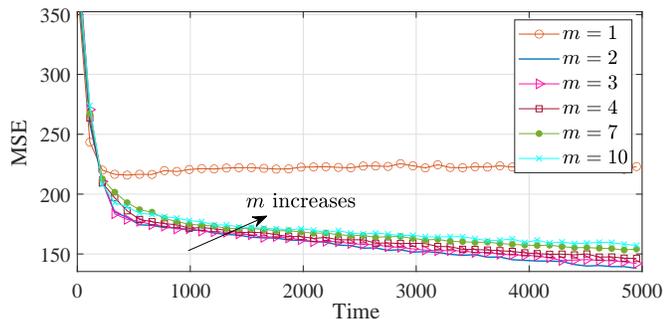}
    \caption{\small{Impact of memory capacity, $m$, on convergence, when CN store the global parameter in erroneous communication, for $U=10$, $|D_u|=100$, $\gamma_0=3$ dB, $\eta = 0.05$, $R = 0.9$ and similar $\alpha$ for all global parameters (equal weighted)}.}
    \label{fig:Global memory on convergence}
   \vspace{-0.5 em}
\end{figure}

When the CN does not have enough storage capacity, instead of storing local parameters of devices, it will store the past $m$ global updates (\ref{CN caches global}).
Fig. \ref{fig:Global memory on convergence} shows the impact of the available memory when CN uses memory to cache the previous $m$ global updates. Here, we considered all the past stored global updates have similar weight ($\alpha = 1/m$). As can be seen, although the performance converges with just one unit of memory ($m = 1$) and no fluctuations are visible, the loss is very high.
Furthermore, we can see that when we employ more than two units of memory ($m > 2$) and equal weight averaging, the system takes longer to converge. The reason for this is because the old global update has a larger gap to the erased local update, and adding additional memory just increases the gap between the value of the erased local update and its projected value using global updates.

\begin{figure}[t]
    \centering
\includegraphics[width=1\columnwidth]{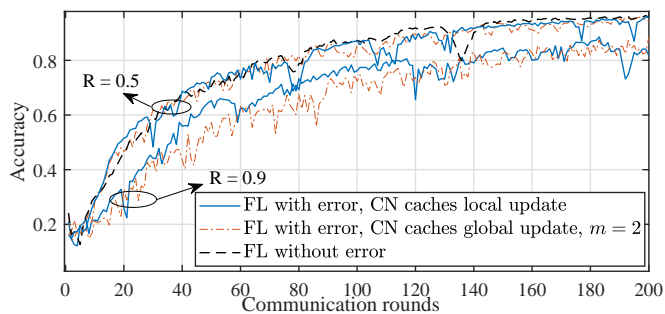}
         \caption{Accuracy vs. communication round for the FL algorithm with MNIST digits dataset, when $U=8$, $\gamma_0=0$ dB, and long packet communications.}
         \label{fig:Accuracy FL}
\end{figure}

To investigate the performance of proposed FL schemes in long packet communication, we consider image classification and apply the FL to train a neural network using a highly non-$\mathrm{i.i.d.}$ dataset.  In order to compute the PER for long-packet communication across a fading channel, given $R$ and the transmitted SNR $\gamma_0$, we use $(\ref{error long packet})$  and check to see if instantaneous received SNR is above the SNR threshold required for error-free communication.
We consider the MNIST digits dataset, which consists of handwritten images of each number $0$ to $7$. We considered eight IoT devices ($U=8$), each of them with $1000$ images of one of the numbers between $0$ and $7$. We employed a MATLAB parallel pool with eight workers and allocated $70\%$ of the dataset to training, $15\%$ to test and $15\%$ to validation. We utilised a CNN with two $5 \times 5$ convolution layers activated using ReLu and a final softmax output layer for each user \cite{mcmahan2017communication}. Fig. \ref{fig:Accuracy FL} shows the accuracy of FL schemes for error-free and erroneous communication for two distinct coding rate $R$. As can be seen for $\gamma_0 = 0$ dB, applying low coding rate results in better overall performance. One can see that erroneous communication with $R = 0.5$ have similar performance to error-free communication and the performance will reaches $90\%$ in less than $100$ communication rounds. However, when we increase the coding rate to $R = 0.9$, after $200$ communication rounds the accuracy will be around $84\%$.
%
%
%
%
%
%
%
\section{Conclusion}
We investigated the performance of the federated learning algorithm in the presence of communication errors and studied the impact of coding rate and block length on the accuracy and convergence. We modeled the communication channels as packet erasure channels, with block length, coding rate, and SNR determining the erasure probability. We proposed two schemes to improve the performance of FL under erroneous communications. We demonstrated the effect of coding rate on the convergence of FL for both short packet and long packet communications considering erroneous transmission. It has been demonstrated that a single memory unit has a significant effect on the performance of FL. While the communication errors are deleterious to the reliability of the packets, the effect can be easily compensated by reusing past local or global parameters, in case of communication errors. This is of significant importance for mIoT systems, as one can relax the reliability requirement, and still achieve the desired level of accuracy withing the required time.  

\bibliographystyle{IEEEtran}
\bibliography{IEEEabrv,BIB}

\end{document}